\documentclass[aps, twocolumn, groupedaddress,superscriptaddress, showpacs]{revtex4-1}
\usepackage{graphicx}
\usepackage{setspace,transparent}
\usepackage{color}
\usepackage{fontenc,subcaption,ragged2e}
\usepackage{mathtools,amssymb,nicefrac}
\usepackage{soul}
\usepackage[utf8]{inputenc}

\DeclareCaptionJustification{justified}{\justifying}
\begin{document}

\title{Spontaneous repulsion in the $A+B\to0$ reaction on coupled networks}

\author{Filippos~Lazaridis}
\thanks{F.~L.~and B.~G.~contributed equally to this paper.}
\affiliation{Center for Complex Systems and Department of Physics, University of Thessaloniki - 54124 Thessaloniki, Greece}
\author{Bnaya~Gross}
\email{Corresponding author: bnaya.gross@gmail.com}
\affiliation{Department of Physics, Bar-Ilan University, 52900 Ramat-Gan, Israel}
\author{Michael~Maragakis}
\affiliation{Center for Complex Systems and Department of Physics, University of Thessaloniki - 54124 Thessaloniki, Greece}
\author{Panos~Argyrakis}
\affiliation{Center for Complex Systems and Department of Physics, University of Thessaloniki - 54124 Thessaloniki, Greece}
\author{Ivan~Bonamassa}
\email{Corresponding author: ivan.bms.2011@gmail.com}
\affiliation{Department of Physics, Bar-Ilan University, 52900 Ramat-Gan, Israel}
\author{Shlomo~Havlin}
\affiliation{Department of Physics, Bar-Ilan University, 52900 Ramat-Gan, Israel}
\affiliation{Institute of Innovative Research, Tokyo Institute of Technology, Midori-ku, Yokohama, Japan 226-8503}
\author{Reuven~Cohen}
\affiliation{Department of Mathematics, Bar-Ilan University, 52900 Ramat-Gan, Israel}

\date{\today}

\begin{abstract}
We study the transient dynamics of an $A+B \rightarrow 0$ process on a pair of randomly coupled networks, where reactants are initially separated.
We find that, for sufficiently small fractions $q$ of cross-couplings, the concentration of $A$ (or $B$) particles decays linearly in a first stage and crosses over to a second linear decrease at a mixing time $t_x$.
By numerical and analytical arguments, we show that for symmetric and homogeneous structures $t_x\propto(\nicefrac{\langle k \rangle}{q})\log(\nicefrac{\langle k \rangle}{q})$ where $\langle k \rangle$ is the mean degree of both networks. 
Being this behavior in marked contrast with a purely diffusive process---where the mixing time would go simply like $\langle k\rangle/q$---we identify the logarithmic slowing down in $t_x$ to be the result of a novel spontaneous mechanism of {\em repulsion} between the reactants $A$ and $B$ due to the interactions taking place at the networks' interface. 
We show numerically how this spontaneous repulsion effect depends on the topology of the underlying networks. 
\end{abstract}


\maketitle
Nearly one hundred years ago, Marian von Smoluchowski introduced a mathematical model to describe coagulation phenomena in terms of diffusion-controlled reaction processes~\cite{Smo917}. 
Despite its apparent simplicity, the kinetics of this model was found to yield a wealth of intriguing phenomena, whose analysis have widely enriched our understanding of pattern formation in chemical compounds, biological systems~\cite{Murray,Kaneko} and elsewhere.\\
\indent
From the Statistical Mechanics perspective, reaction-diffusion (RD) processes represent a fertile groundwork where to analyze the emergence of spontaneous mechanism by starting from microscopic rules~\cite{kampen1992}.
Most studies in this direction aimed to unveil the effect that dynamical correlations and geometrical (or topological) constraints of the underlying structures have on the spatiotemporal evolution of the reactants' concentrations~\cite{ShlomoBen}.\\
\indent 
The $A+B\to0$ process, in particular, is known to exhibit anomalous kinetics on low dimensional and fractal geometries, where density fluctuations yield the formation of {\em self}-{\em segregation} domains composed of particles of the same type~\cite{OZ,Wilczek,Redner,Front}. 
These phenomena result in a drastic slowing down in the rate of the reactions, forcing the system in a long-lived non-equilibrium state with a sub-linear decay in the density of the surviving particles. 
Since this type of process grasps the essential kinetics featured by the spreading of pathogen-antipathogen agents~\cite{Castellano2009,Vespignani}, or underlying the pattern-formation of diverse chemical reaction~\cite{Galfi,Ovchinnikov,Taitelbaum}, the appearance of sub-diffusive dynamics may become inefficient for practical applications, which typically aim in fast mixing regimes. With this goal, it was later proved that the adoption of L\'evy processes indeed washes out segregation phenomena, leading to super-diffusive dynamics~\cite{Levy1996}.\\
\indent 
After this ``classical'' period, the study of RD processes has experienced a relevant boost with the inception of network theory as a new field for characterizing the structures underlying real-world complex systems~\cite{Newman, Reuvenbook}. 
In fact, besides the focus on networks' topological properties, a mainstream has been (and still is) to understand the interplay between their structure and the dynamics of processes taking place on them~\cite{Sbrocca}.\\
\indent 
Consistently with the scenario observed in other models~\cite{Calda,collective}, complex networks significantly influence the collective properties of RD processes~\cite{pastorsatorras2015}.
Numerical~\cite{gallos1,gallos2} and theoretical~\cite{catanzaro} results have showed that the small-world property of these substrates mitigates the local fluctuations in the particles' density, facilitating their reactions. 
Notably, on scale-free (SF) networks (i.e.~random graphs with connectivity distribution $P(k)\sim k^{-\gamma}$ and $2<\gamma\leq3$) the kinetics of the $A+B\to0$ process exhibit jamming effects at early stages and then super-diffusive behaviors~\cite{weber}, with the latter becoming stronger as the network heterogeneity increases. 
Homogeneous structures---like random regular (RR), Erd\H{o}s-R\'enyi (ER) or scale-rich (SR, i.e.~$P(k)\sim k^{-\gamma}$ and $\gamma>3$) networks---result instead in a linear decay of the density, in accordance with the mean-field predictions~\cite{gallosRev,catanzaroRev}.\\
\indent 
Though the portrait of RD processes on isolated structures is nowadays clear, not much is known regarding their dynamical behaviors on multilayer networks~\cite{gomez,garas}. 
The influence that their mesoscopic organization has on the collective behaviors of processes acting on them is attracting significant interest~\cite{havlin,boccaletti,Kivela,ginestra,garasbook}, and has already produced interesting results~\cite{interdep1,interdep2,bornholdt17}. 
Increasing evidence, in fact, is showing that the existence of multiple layers, joined with the possibility of modeling different types of cross-system interactions, results in novel collective behaviors whose analysis is still in its infancy~\cite{charoSA,arenas17,micha17}. Following this mainstream, we study here the $A+B\to0$ dynamics on a pair of randomly coupled networks with initially separated reactants, where we find a novel spontaneous dynamical phenomenon.\\
\indent 
Our results show that, for sufficiently small fractions $q$ of the cross-couplings between the layers, the concentration of both reactants decays as $\rho(t)\sim C_1/t$ for a long transient, and then crosses over to a second linear regime where $\rho(t)\sim C_2/t$ (with $C_2\ll C_1$) at a mixing time $t_x$. 
We interpret the initial transient ($t<t_x$) as the unmixed regime, where the reactions between $A$ and $B$ particles take place mainly at the boundaries between the networks (i.e.~at the interconnected nodes). 
At larger times ($t\approx t_x$), the reactants penetrate more and more the two layers and start to react everywhere in the system.
After this fast mixing stage ($t>t_x)$, the remaining particles are uniformly distributed in the system and their dynamics is driven essentially by diffusion. 
We find that the {\em mixing time} $t_{x}$ depends on the ratio $q/\langle k\rangle$, where $\langle k\rangle$ is the average degree of the networks, according to the formula
\begin{equation}\label{eq:tx}
t_{x}\propto \frac{\langle k \rangle}{q}\log\left(\frac{\langle k \rangle}{q}\right),
\end{equation} 
\noindent 
that we derive analytically for RR graphs and verified by extensive simulations on diverse synthetic networks. 
Since Eq.~\eqref{eq:tx} is in marked contrast with a purely diffusive process, where $t_x$ would simply scale as $\langle k\rangle/q$, we interpret the logarithmic factor as the reflection of a {\em spontaneous} {\em repulsion} mechanism between reactants due to the reactions taking place at the boundary between the networks. 
We find that this mechanism becomes stronger with increasing heterogeneity of the underlying structures, in which case Eq.~\eqref{eq:tx} holds only approximately.\\
\indent 
The paper is organized as follows. 
We derive analytically Eq.~\eqref{eq:tx} for RR graphs, that we verify against extensive simulations on uncorrelated configuration model (UCM) networks. 
After investigating the effects that the underlying topology has on $t_x$ and on the dynamical regimes observed, we give our conclusions.\vspace{+0.2cm}

\begin{figure}
	\includegraphics[width=.48\textwidth]{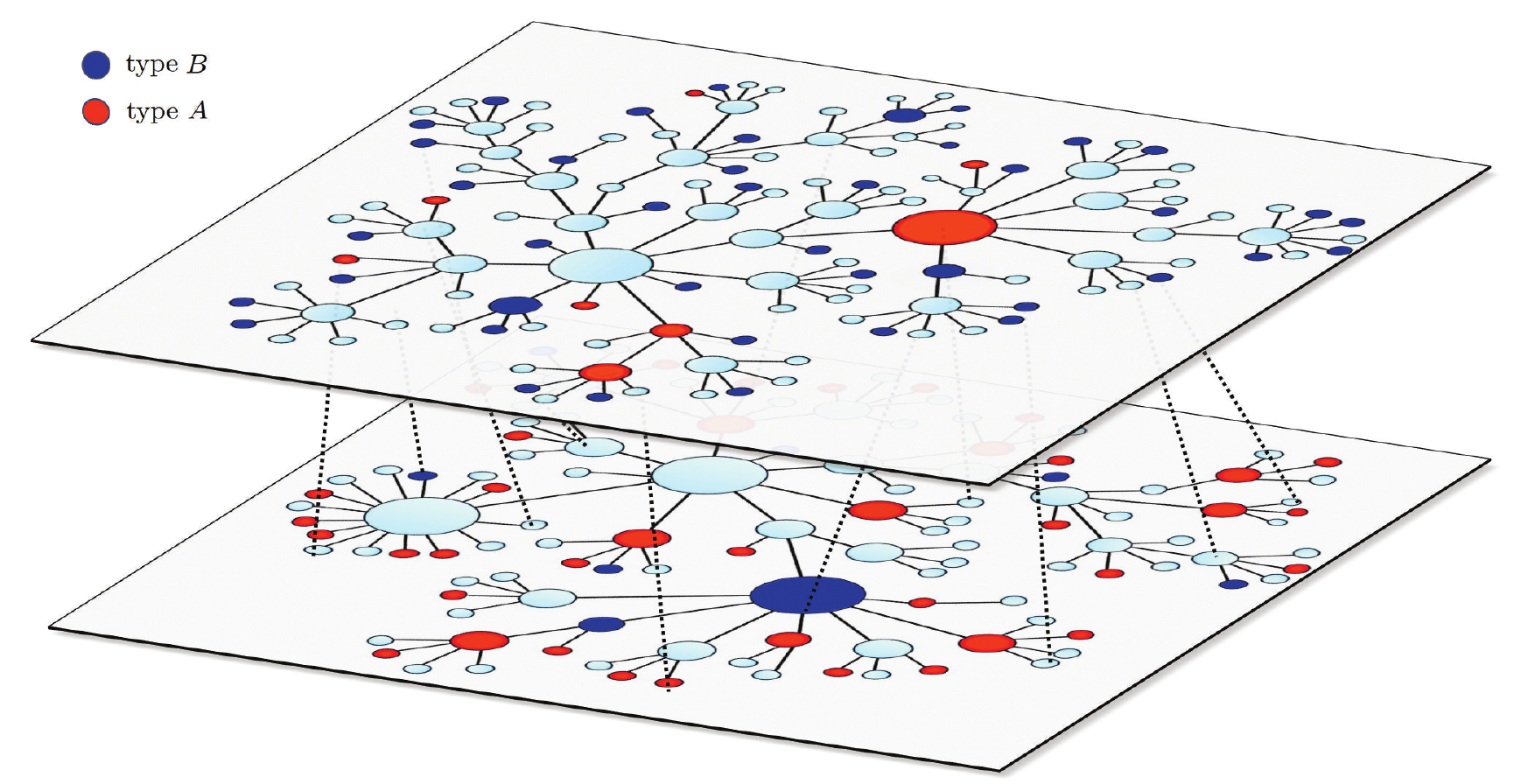}
		\centering
		\caption[]{(Color online) Illustration of the model. Particles of type $A$ (red) and $B$ (blue) are let to diffuse and react on the top of an interconnected network composed of two layers, each having a given degree distribution and a fraction $q$ of interconnected nodes. Different nodes' size pictorially represent degrees' heterogeneity.}\label{fig:RDmodel}
\end{figure}

\paragraph*{\textbf{\em I. Analytic approach.}}
We consider two configuration model networks~\cite{Bollobas}, composed by the same number of nodes and same structural properties, a condition that we will refer hereafter as the ``symmetry'' of the interconnected network.
Let us further assume the two layers are coupled by means of undirected interlinks, placed at random between a fraction $q\in[0,1]$ of couples of nodes belonging to different layers (Fig.~\ref{fig:RDmodel}). 
Two populations of $A$ and $B$ reactants are then randomly distributed with initially separated concentrations, so that all the particles of the same type are placed on the same layer.
For simplicity we assume that the initial concentrations of reactants are equal.
To track the evolution of each populations, let us denote by $\rho_1$ and $\rho_2$ the concentration of $A$ particles in network 1 and 2, respectively; similarly, let $\mu_1$ and $\mu_2$ be the concentration of $B$ particles in network 1 and 2, respectively. 
Having assumed equal initial conditions, we have $\rho_1(0)=\mu_2(0)=\rho_0$ and $\rho_2(0)=\mu_1(0)=0$. 
Moreover, by symmetry, $\rho_1(t)=\mu_2(t)$ and $\rho_2(t)=\mu_1(t)$ for all $t$. 
It is worth to notice here that, whilst this symmetric condition certainly holds in the case of coupled layers with the same or mildly different topological features (say RR and ER layers, or two ER networks with different average degrees), it will in general require some adjustments for layers with different structures.\\
\indent 
To further simplify the analysis, let us assume that the networks underlying the RD process have homogeneous topologies, so that the average degree $\langle k\rangle$ of nodes is the only characteristic parameter of the structure. 
In this case, disregarding any dynamical effect due to the topological fluctuations, we assume that the overall behavior is captured by the average densities $\rho_1$ and $\rho_2$.
We can then describe the rate of change of the concentrations $\rho_1$ and $\rho_2$ which, by symmetry, is given by
\begin{align}
{\dot\rho}_1&\,=-\tilde q\rho_1 +\tilde q\rho_2-\rho_1\rho_2\;,\label{eq:rho1}\\
{\dot\rho}_2&\,=-\tilde q\rho_2 +\tilde q\rho_1-\rho_1\rho_2\;,\label{eq:rho2}
\end{align} 
where the first two terms in both lines are due to diffusion, and the last term is due to reaction.
The effective diffusion rate $\tilde q=q/\langle k\rangle$ is the probability at each node to move to the other network.\\
\indent 
Subtracting Eq.~\eqref{eq:rho1} from Eq.~\eqref{eq:rho2}, one obtains ${\dot\rho}_1-{\dot\rho}_2=-2\tilde q(\rho_1-\rho_2)$ whose solution is $\rho_1-\rho_2=\rho_0 e^{-2\tilde q t}$. 
Inserting this expression into, say Eq.~\eqref{eq:rho1}, leads to 
\begin{equation}
\label{eq:final}
\dot\rho_1=\rho_0\big(\rho_1-\tilde q\big)e^{-2\tilde q t}-\rho_1 ^2\;.
\end{equation}
which is a Riccati differential equation. Being the densities now decoupled, we can focus hereafter only on the solutions of one of them, dropping dumb labels. 
Solving numerically Eq.~\eqref{eq:final} will give us the theoretical predictions for the evolution of the reactants' concentrations.\\
\begin{figure*}
\hspace{+17.1cm}\llap{{\!\!\!\!\includegraphics[width=.30\textwidth]{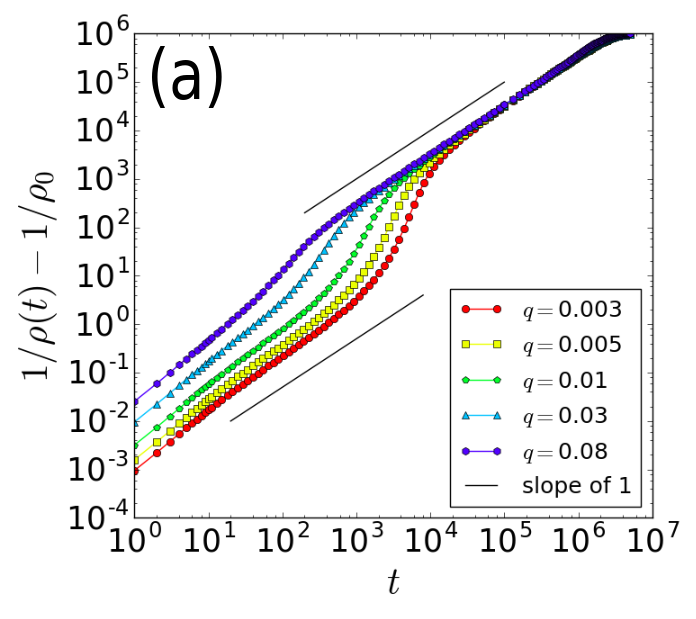}
\includegraphics[width=.32\textwidth]{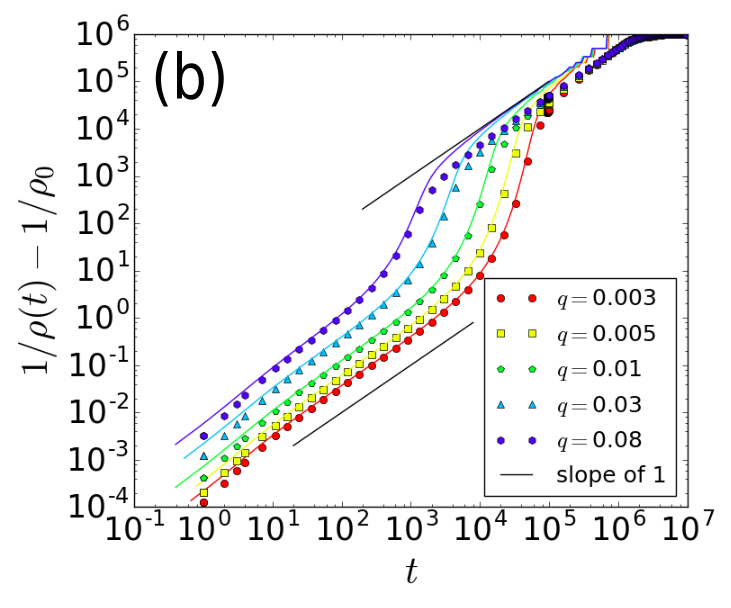}
{\hspace{-3.1cm}\llap{\raisebox{2.4cm}{\includegraphics[width=.085\textwidth]{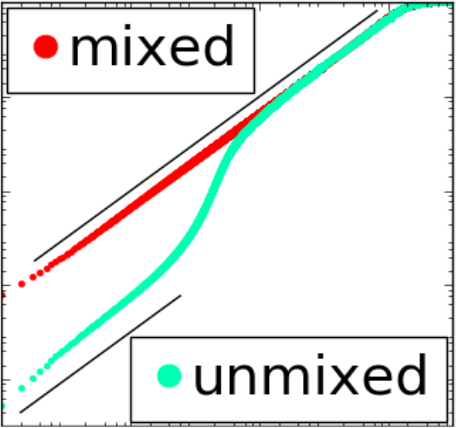}}}}
\hspace{+2.8cm}\llap{}{\includegraphics[width=.32\textwidth]{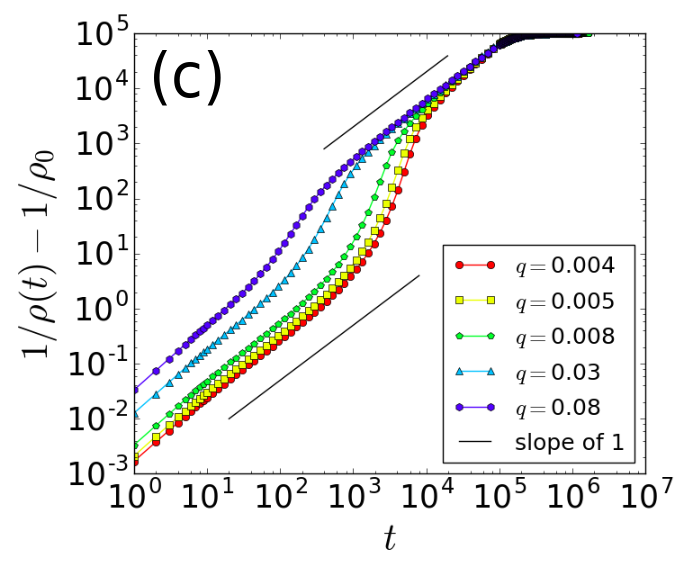}}}}\\
\raisebox{0.08cm}{\includegraphics[width=.315\textwidth]{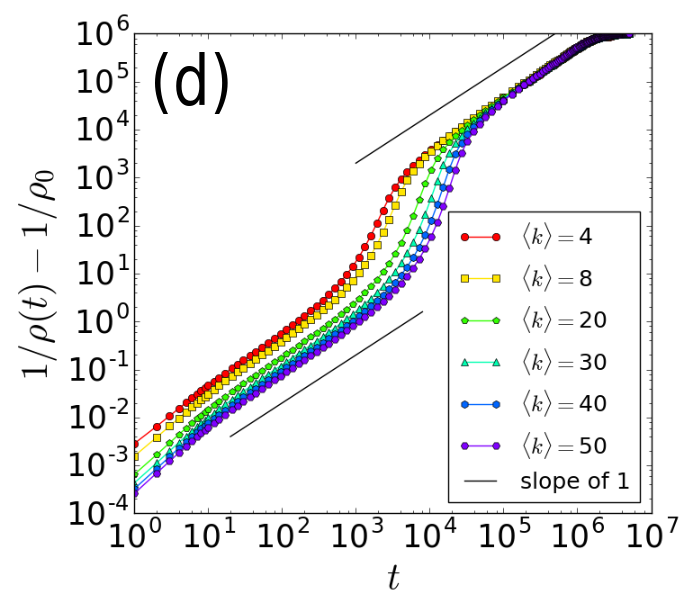}}
\hspace{-0.3cm}\includegraphics[width=.32\textwidth]{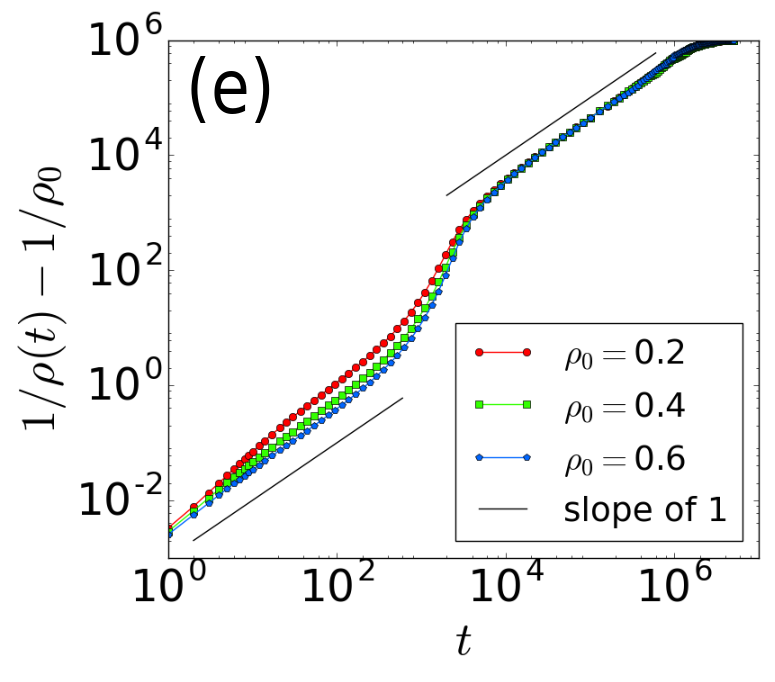}
\raisebox{0.15cm}{\includegraphics[width=.32\textwidth]{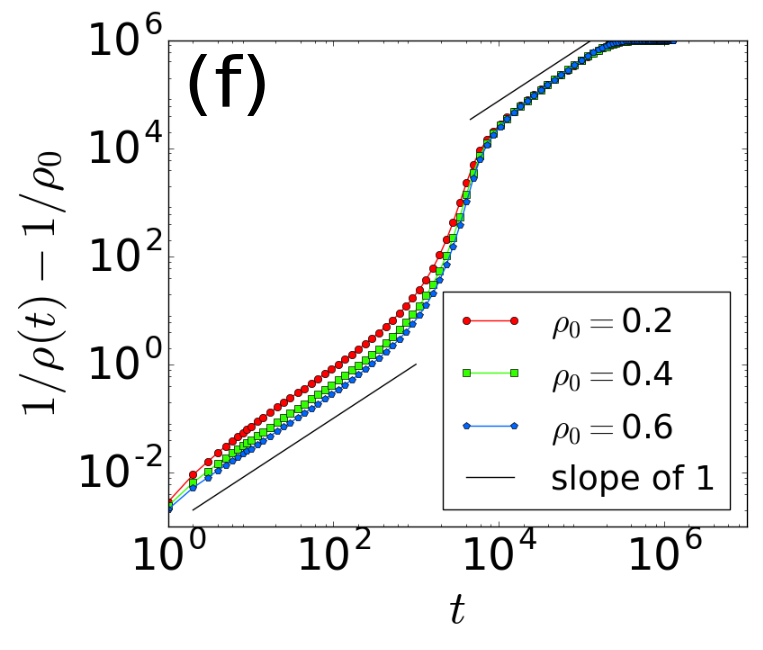}}
\caption{(Color online) Time evolution of the inverse density in log-log scale for the $A+B\to0$ process on a pair of randomly coupled networks of equal number of nodes $N=10^6$ and the initial concentrations $\rho_0=0.4$. Different markers in the figures above correspond to different choices of the varying parameters considered.
\textbf{(a)} RR networks with $\langle k \rangle = 4$ and decreasing values of $q$. 
\textbf{(b)} ER networks with $\langle k \rangle = 40$ and decreasing values of $q$. Simulation's data are represented by dots, whilst full lines are obtained by integrating numerically Eq.~\eqref{eq:final}. The inset shows the difference between mixed and unmixed initial conditions for $\langle k \rangle=4$ and $q=0.008$.
\textbf{(c)} SF networks with $\gamma = 3.5$ and decreasing values of $q$.
\textbf{(d)} ER networks with fixed fraction $q = 0.008$, and increasing values of $ \langle k \rangle$. 
\textbf{(e)} ER networks with $q=0.008$, $\langle k \rangle=4$ and different initial concentrations.
\textbf{(f)} SF networks with $q=0.008$, $m=2$ and $\gamma=2.5$, with different initial concentrations. Both plots in (e) and (f) have been averaged over $1000$ realizations.} {\label{Figure01}}
\end{figure*}
\indent 
Let us now notice that Eq.~\eqref{eq:final} is characterized by two distinct regimes: a {\em reaction}-{\em dominated} regime, where $\dot{\rho}=-\rho^2$ with solution $1/\rho-c=t$ which can be understood as the asymptotic behavior of the system, and a {\em diffusion}-{\em limited} regime where the exponentially decaying terms are dominant over the reaction one. 
By comparing these two regimes, we obtain an equation for the quasi-stationary behavior, namely
\begin{equation}\label{eq:quasi}
\rho ^2 \approx \rho_0\left|\rho-\tilde q \right|e^{-2\tilde q t}\;.
\end{equation}
\noindent
Eq.~\eqref{eq:quasi} can be adopted in order to define the mixing time $t_x$, that is the characteristic time it takes for the $A$ and $B$ reactants to mix and react like in a single network. 
To this aim, we make the assumption that the second summand on the right hand side of Eq.~\eqref{eq:quasi} is dominant, i.e.~$\tilde q\gg\left.\!\rho\,\right|_{t=t_x}$~\cite{comment}.
In this approximation, we get
\begin{equation}\label{eq:approx}
\log\rho\approx -\tilde q t_x\log(\tilde q\rho_0)\;,
\end{equation}
which, in combination with the solution of the reaction-limited regime, gives to the leading order Eq.~\eqref{eq:tx}.\vspace{+0.2cm}

\paragraph*{\textbf{\em II. Simulations and results}}
To simulate the $A+B\to0$ process on two randomly coupled networks, we have first generated two equal synthetic networks composed of $N=10^6$ nodes where the reactants will be initially distributed. 
Once the networks are prepared, we randomly place the cross-couplings between them with probability $q\in[0,1]$ by quenching the labels of interconnected nodes.
To avoid any dynamical effect due to degree-degree correlations, we have constructed two random networks according to the UCM.
Each network is hence assigned with a specific degree sequence having the desired connectivity distribution for the structure, with lower and structural cut-off given by $m\leq k_i\leq N^{1/2}$, where $m\geq1$ is the minimum degree of each node \cite{UCM}.\\
\indent 
Two population of reacting ($A$ and $B$) species are then randomly distributed on the interconnected network with initially separated concentrations, meaning that all $A$ particles are placed on nodes of one layer, and all $B$ particles on nodes of the other. 
Following our symmetric choice for the system, we fix the initial densities of reactants to be the same, so that $\rho_1(0)=\mu_2(0)\equiv \rho_0$.\\
\indent 
Particles diffuse in the system by performing independent random walks, where hops are allowed only to nearest neighbor nodes.
Being interested in studying an $A+B\to0$ process, we further assume that reactants of the same species do not interact with each other once they occupy the same site simultaneously, i.e.~we adopt a {\em bosonic} version of the dynamics~\cite{bosonic1,bosonic3}.
A reaction occurs whenever an $A$ and a $B$ particle occupy the same site, in which case both reactants generate an inert species and are then removed from the system.
We monitor the time evolution of the concentrations of $A$ and $B$ particles, where the total time advances at each step as $1/(n_A+n_B)$, being $n_A+n_B$ the number of particles currently present in the system. 
Results are then averaged over a set of realizations, whose nominal cardinality is hereafter assumed to be $300$, unless otherwise stated.\\
\indent
In Fig.~\ref{Figure01} we present the inverse particle density as a function of time for decreasing values of $q$ in coupled RR (Fig.~\ref{Figure01}a), ER (Fig.~\ref{Figure01}b) and SF networks (Fig.~\ref{Figure01}c). 
In all the cases, we find that for small enough values of the fractions $q$ of interconnected nodes, three distinct dynamical regimes exist.
One for short times ($t<t_x$), where the particles diffuse inside their own layer and reactions occur mainly at the interface, one for intermediate times ($t\approx t_x$), where particles start to cross and react in the opposite layer, and a third one for long times ($t>t_x$), where eventually the survived reactants are well mixed and the coupled systems behave like a single network.
As shown in the inset of Fig.~\ref{Figure01}b, this kinetic pattern strongly depends on the choice of an initial separation of the reactants, and eventually disappears when the particles concentrations are initially mixed. 
To demonstrate these results, we have compared in Fig.~\ref{Figure01}b the numerical solution obtained from the theory and given by Eq.~\eqref{eq:final} with the data collected from the simulation, in which case we observe an excellent agreement.\\
\indent
For homogeneous structures, we have also tested the effects that the average degree has on the mixing of the system. 
In Fig.~\ref{Figure01}d, it is depicted again the inverse particle density as a function of time for ER networks, only this time with a fixed fraction $q = 0.008$ and increasing values of the mean degree. 
For these substrates, we find that the same repulsion mechanism is obtained by both increasing $\langle k\rangle$ or decreasing $q$, as one might expect since a particle hopping to a node which is cross-coupled to the other layer will diffuse to it with probability $q/\langle k\rangle$. 
Finally we tested the sensitivity of this dynamical scenario with respect to different choices of the initial concentrations $\rho_0$. 
As shown in Fig.~\ref{Figure01}e,f, increasing values of $\rho_0$ slightly affect the dynamical behavior of the system for both ER and SF interconnected networks, mainly influencing the transients in which the system indulges before entering the first diffusive regime.
In particular, higher values of the densities make the particles in the two systems to experience earlier the repulsive effects.\\
\indent 
At this point, it worth to notice that the phenomenology described so far partially agrees with the results presented earlier in Ref.~\cite{garas} by A.~Garas, where the same system was investigated from the more general perspective of different strategies of cross-systems interactions.
However, by contrast with the conclusions drown in Ref.~\cite{garas}, here we have shown that in the limit of low enough fractions of interconnected nodes, an initial separation of reactants generally leads to a novel spontaneous mechanism of repulsion among the reactants which, at the best of our knowledge, has been so far overlooked.
\begin{figure}[h!]
\includegraphics[width=0.41\textwidth]{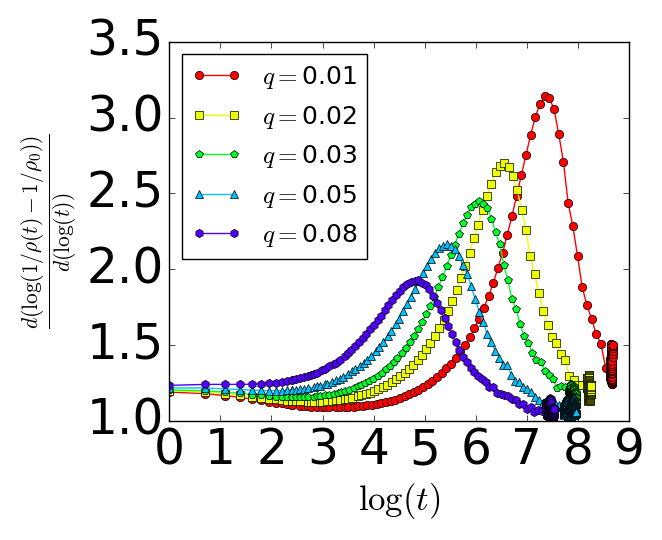}
\caption{(Color online) Time evolution for the rates of particle's annihilation. 
Results are obtained for ER networks with $N = 10^6$, $\rho_0 = 0.4$, $\langle k \rangle = 4$, and decreasing values of $q$ represented by different markers. 
}
{\label{Figure02}}
\end{figure} 
\begin{figure}\centering
	{\hspace{0cm}\includegraphics[height=0.8\linewidth,width=.44\textwidth]{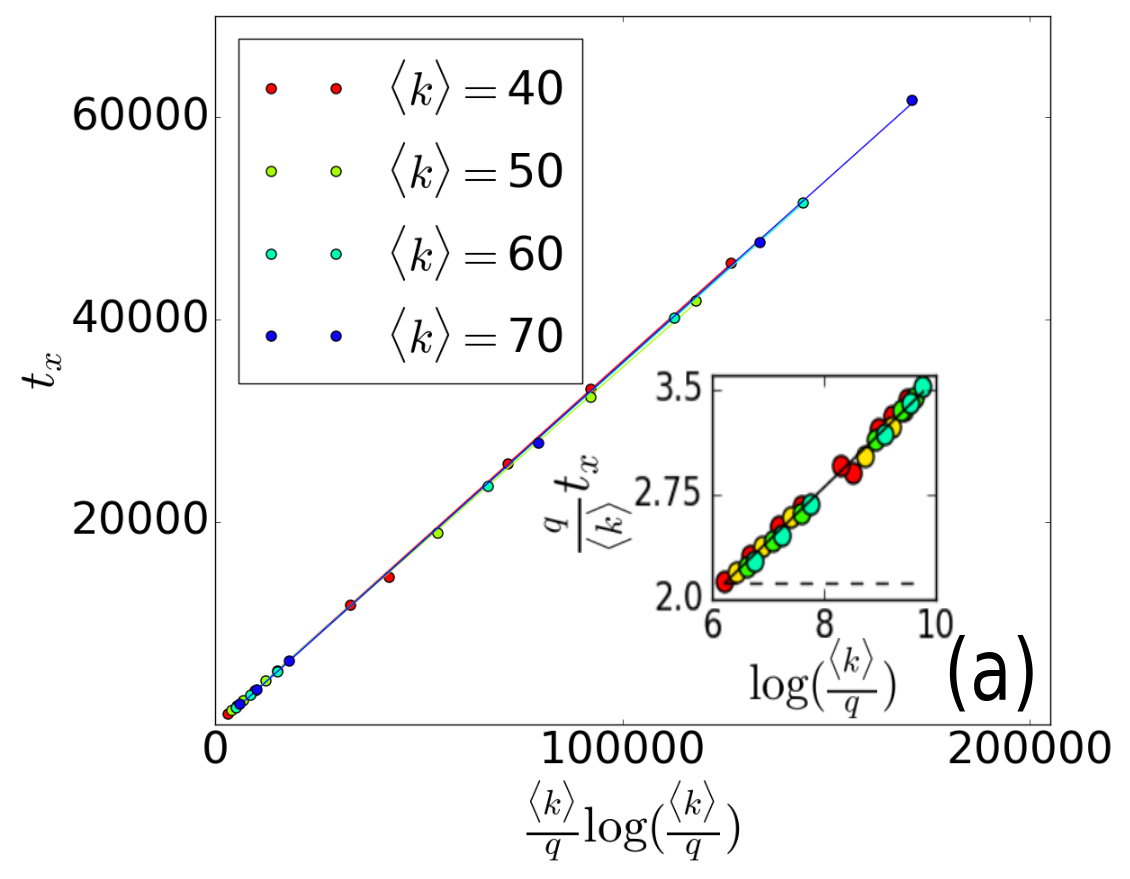}}\\
	\includegraphics[width=.44\textwidth]{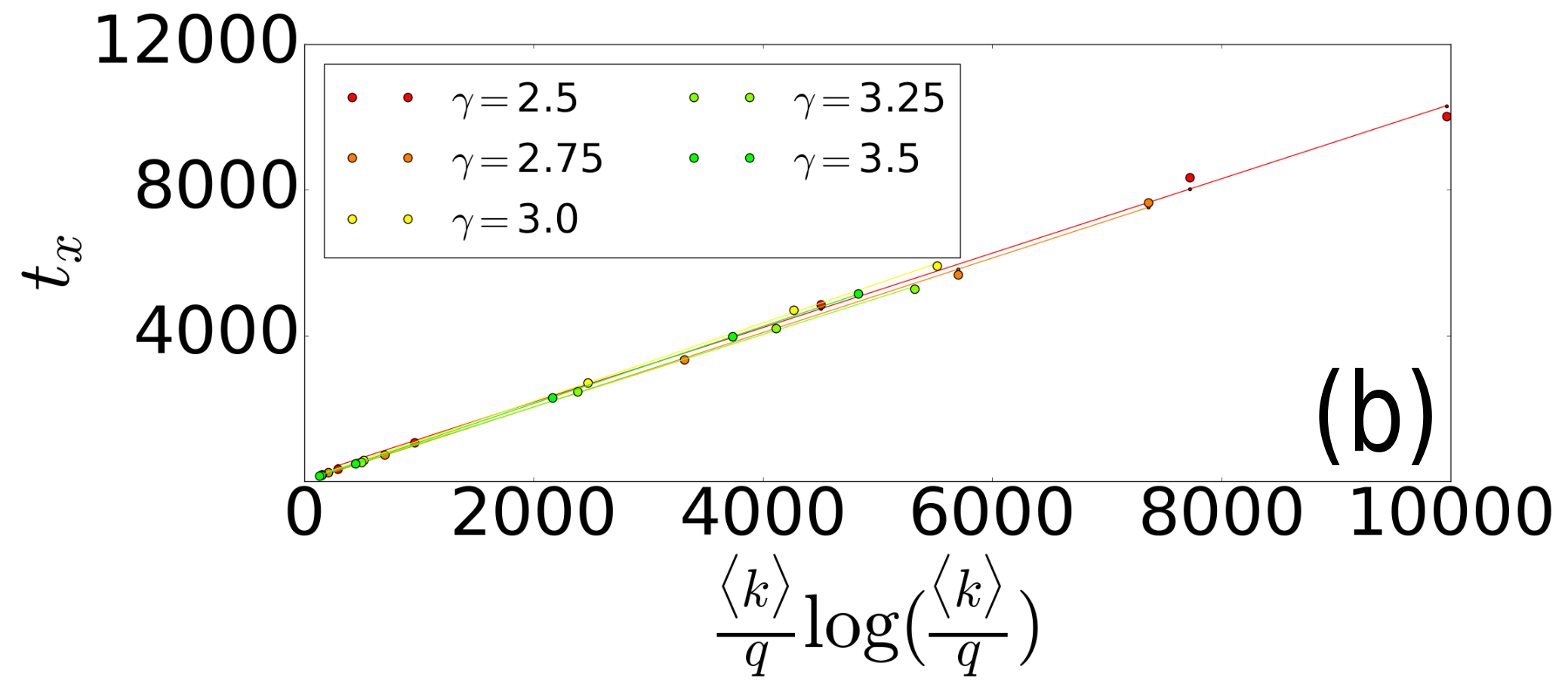}
		\centering
		\caption[]{(Color online) Data collapse of the mixing time.
		\textbf{(a)} Results for ER networks with $N=10^6$ and different values of $\langle k \rangle$. The inset demonstrates the logarithmic dependency of $t_x$ on $\langle k\rangle /q$ as observed in the same data (circles) and predicted by the theory (full line).
		\textbf{(b)} Results for SF and SR networks with $N=10^6$, $m=2$, and increasing values of the $\gamma$ exponent.}\label{crossover_tc}
\end{figure}
\indent 
Next, we investigate the effects that different values of the effective transmission probability $q/\langle k\rangle$ has on the mixing properties of the system, so to verify the accuracy of the approximations adopted to derive the relation \eqref{eq:tx} for $t_x$.
To evaluate this quantity, we have plot the logarithmic derivative of the $y$-axis in Fig.~\ref{Figure01}, and searched for the maxima of the corresponding curves (see Fig.~\ref{Figure02}). The time at which a maximum is reached defines the mixing time $t_x$, sharply marking the dynamical crossover between the two regimes. 
As shown in Fig.~\ref{crossover_tc}a, we find that the time $t_x$ for the $A+B\to0$ process on ER (and equivalently, though not shown, RR) networks depends on $q/\langle k \rangle$ as predicted by Eq.~\eqref{eq:tx}. 
To further support this behavior, in the inset of Fig. \ref{crossover_tc}a we have validated the logarithmic dependency of $\frac{q}{\langle k \rangle}t_x$ due to the repulsion mechanism by comparing the pattern observed with the constant behavior that one would have found in the case of a purely diffusive process.
\begin{figure}
\hspace{-0.95cm}
	\includegraphics[width=.25\textwidth]{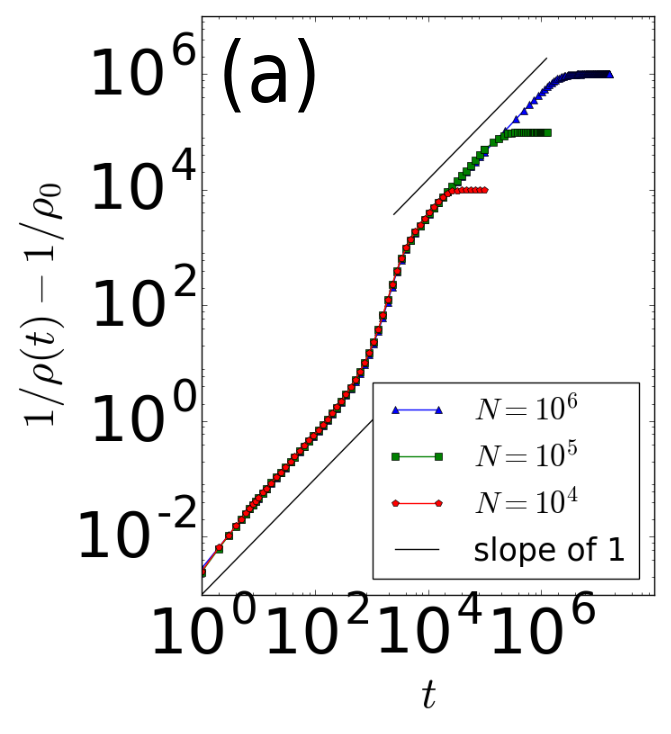}
	\raisebox{0.2cm}{\includegraphics[width=.27\textwidth]{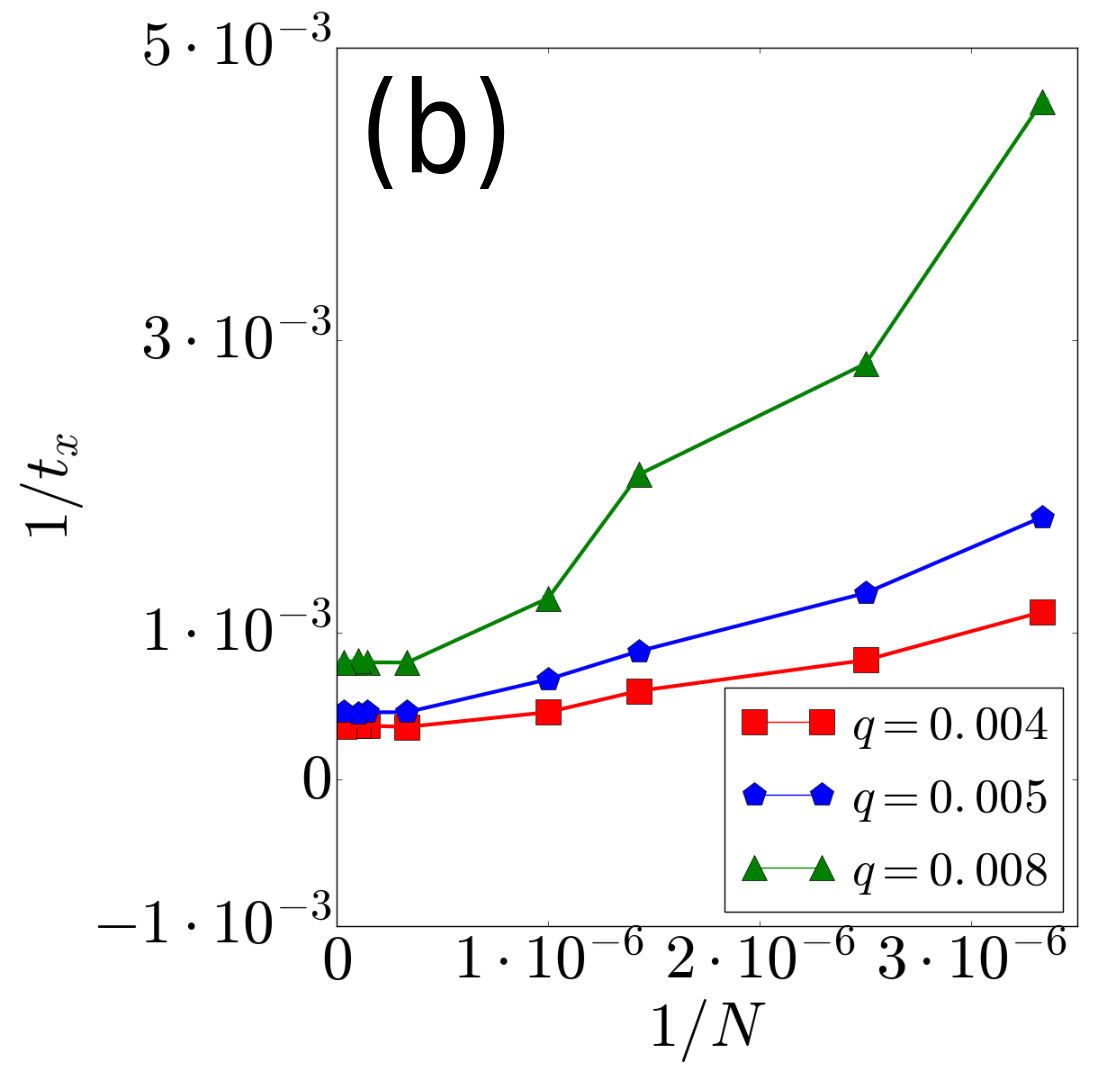}}
		\centering
		\caption[]{(Color online) Finite-size effects. 
		\textbf{(a)} Inverse density vs.~time for ER networks with $\langle k\rangle=4$, $\rho_0=0.4$, $q=0.008$ and different sizes, as specified by the markers. 
		Besides an earlier extinction of reactants (flat curves, order $1/N$) for smaller networks, the mixing time $t_x$ and the kinetic stages remain unaltered. 
		\textbf{(b)} Plot of $1/t_x$ vs.~$1/N$ for SR networks with $\gamma=3.5$, $m=1$ and initial concentration $\rho_0=0.4$. 
		Notice that $t_x$ converges to a finite value for large enough networks.}\label{Finite_size}
\end{figure} By performing the same analysis on heterogeneous (SF and SR) networks, we find that the mixing time $t_x$ surprisingly follows the behavior predicted by the mean-field theory for homogeneous structures (Fig.~\ref{crossover_tc}b), though with slightly less accuracy. 
We trace the origin of this result to the fact that, in our model, low-degree nodes are indeed the most important ones for reactions to occur between the two populations, as they most likely carry the cross-connections. 
In fact, as percolation results would suggest~\cite{SFperc1}, the probability of picking at random a hub is very low in SF or SR networks, making even harder the chance of having hub-to-hub interconnections. In this respect, we expect that degree-based couplings among networks will likely tune the effects of the repulsion mechanism, leading to a faster mixing of the reactants for e.g.~hub-to-hub couplings.\\
\indent 
To complete the analysis of the model for this Rapid Communication, we have tested the response to the system to finite-size effects, by adopting networks with equal topologies, but different number of nodes. 
In particular, for ER networks the three dynamical regimes encountered in Fig.\ref{Figure01}b remain unaltered in their main stages (Fig.~\ref{Finite_size}a), and only exhibit an extinction point at earlier times for networks of decreasing sizes. 
Finally in Fig.~\ref{Finite_size}b we have repeated the test for SR networks, where we have found that the pattern observed in Fig.~\ref{Figure01}c is again mainly unchanged, except that the convergence of the mixing time $t_x$ to its thermodynamic value is monotonic in $N$, enlightening the possible occurrence of finite-size effects in the case of power-law networks of small-size whose study will be performed elsewhere.\vspace{+0.2cm}

\paragraph*{\textbf{\em III. Summary and conclusions}}
In this work we have studied, both numerically and analytically, the dynamics of an $A+B \rightarrow 0$ process on a pair of randomly coupled networks, where reactants are initially separated. 
For small enough values of the fraction $q$ of interconnected nodes between the layers, we have found that the inverse particle density scales linearly at short times and then crosses over to a second linear regime at time $t_x$. 
As the crossover determines the time at which the two population start to extensively mix, we have analyzed the dependence of the mixing time $t_x$ on the effective diffusion rate $\tilde q=q/\langle k\rangle$, unveiling a novel {\em repulsive mechanism} whose spontaneous emergence delays the mixing of the reactants.
We gave numerical evidence that, on randomly coupled synthetic networks, this effect does not show a sensitive dependence on the heterogeneity of the underlying topology, but it is in fact dominated by nodes with low connectivity. 
Whether or not the same behavior will appear on networks with targeted (e.g.~hub-to-hub, or non-hub-to-hub) interconnections~\cite{Aguirre} or on more realistic structure having e.g.~a spatial embedding or degree-degree correlations, remains an intriguing question calling for further investigations. 
Moreover, since the diffusion-controlled annihilation process adopted in this work can be considered as an archetypal model for reaction kinetics~\cite{Redner2}, we believe that these results will inspire the investigation of the effects that the initial distribution of reactants, together with the mesoscopic architecture of the interconnected network, will have on the pattern formation in more elaborated and realistic spreading models~\cite{Castellano2009,agliari}, enlarging in this way our understanding of the interplay between structure and dynamics.

\section*{Acknowledgments}
Results presented in this work have been produced using the European Grid Infrastructure (EGI) through the National Grid Infrastructures $NGI \_ GRNET$ (HellasGrid) as part of the SEE Virtual Organisation. This research was supported by European Commission FP$7$-FET project Multiplex \#~$317532$. S.H.~acknowledges the ISF, ONR Global, the Israel-Italian collaborative project NECST, Japan Science Foundation, BSF-NSF, and DTRA (Grant \#~HDTRA-1-10-1-0014), together with the Israeli Ministry of Science, Technology and Space (MOST, grant \#3-12072) for financial support.


\begin{thebibliography}{}
\bibitem{comment} After the crossover, the reactants become (so to say) oblivious of the “mesoscopic” structure of the interconnected network. This means that the fraction of interconnected nodes must be much larger then the density of the survived particles, which justifies the condition $\tilde{q}\gg \rho|_{t=t_x}$ adopted in deriving Eq.~\eqref{eq:approx}.
\end{thebibliography}
\end{document}